\documentstyle[12pt,qqa4lart]{article}
\input tcilatex

\QQQ{Language}{
American English
}

\begin{document}

\author{Lu-Ming Duan and Guang-Can Guo\thanks{%
E-mail: gcguo@sunlx06.nsc.ustc.edu.cn} \\
Department of Physics, University of Science \\
and Technology of China, Hefei, 230026, P.R.China}
\title{Linearly-independent quantum states can be cloned }
\date{}
\maketitle

\begin{abstract}
\baselineskip 20pt A fundamental question in quantum mechanics is, whether
it is possible to replicate an arbitrary unknown quantum state. Then famous
quantum no-cloning theorem [Nature 299, 802 (1982)] says no to the question.
But it leaves open the following question: If the state is not arbitrary,
but secretly chosen from a certain set $\$=\left\{ \left| \Psi
_1\right\rangle ,\left| \Psi _2\right\rangle ,\cdots ,\left| \Psi
_n\right\rangle \right\} $, whether is the cloning possible? This question
is of great practical significance because of its applications in quantum
information theory. If the states $\left| \Psi _1\right\rangle $, $\left|
\Psi _2\right\rangle $, $\cdots ,$ and $\left| \Psi _n\right\rangle $ are
linearly-dependent, similar to the proof of the no-cloning theorem, the
linearity of quantum mechanics forbids such replication. In this report, we
show that, if the states $\left| \Psi _1\right\rangle $, $\left| \Psi
_2\right\rangle $, $\cdots ,$ and $\left| \Psi _n\right\rangle $ are
linearly-independent, they do can be cloned by a unitary-reduction process.\\

\end{abstract}

\baselineskip 20pt\newpage\ With the development of quantum information
theory [1], cloning of quantum states arouses great interests. The
no-cloning theorem [2] shows that, cloning of an arbitrary unknown state is
impossible. However, in quantum information theory, such as in quantum
cryptography [3-5], the state is usually not completely arbitrary, but
secretly chosen from a certain set $\$=\left\{ \left| \Psi _1\right\rangle
,\left| \Psi _2\right\rangle ,\cdots ,\left| \Psi _n\right\rangle \right\} $%
. So a practical question is, whether the state can be cloned in this
circumstance. If the cloning machine is limited to the unitary evolution,
Refs. [6] and [7] show the impossibility of cloning two non-orthogonal
states. Ref. [8] has extended this result and proven that, two non-commuting
mixed states can not be broadcast onto two separate quantum systems by a
unitary process, even when the states need only be reproduced marginally.
Nevertheless, in quantum mechanics, apart from the unitary evolution, the
reduction process also plays an essential role, such as in the preparation
[9,10] or in the communication [11,12] of quantum states. A nartural
question is thus, whether two non-orthogonal states can be cloned by a
unitary evolution together with a reduction process. Ref. [13] says yes to
the question. The positive answer causes one further to think, whether
multi-states, which belong to a certain finite set, all can be cloned by the
unitary-reduction process. The answer is obviously negative since either the
unitary or the reduction process is linear. The linearity of quantum
mechanics forbids replication of linearly-dependent quantum states.
Surprisingly, here we further prove that, if the states are
linearly-independent, they do can be cloned by a unitary-reduction process.
The result, posed formally, is the following theorem.

{\it Theorem.} The $n$ states $\left| \Psi _1\right\rangle $, $\left| \Psi
_2\right\rangle $, $\cdots ,$ and $\left| \Psi _n\right\rangle $ can be
cloned by the same cloning machine if and only if they are
linearly-independent.

{\it Proof.} If the states $\left| \Psi _1\right\rangle $, $\left| \Psi
_2\right\rangle $, $\cdots $, and $\left| \Psi _n\right\rangle $ are
linearly-dependent, say, $\left| \Psi _n\right\rangle =\stackrel{n}{%
\stackunder{i=1}{\sum }}c_i\left| \Psi _i\right\rangle $, similar to the
proof of the no-cloning theorem [2], the linearity of quantum mechanics
means that, if a cloning machine can replicate $\left| \Psi _1\right\rangle $%
, $\left| \Psi _2\right\rangle $, $\cdots $, and $\left| \Psi
_{n-1}\right\rangle $, it can not replicate $\left| \Psi _n\right\rangle $
at the same time by any way. So our main task is to prove the converse, that
is, we need to show, if $\left| \Psi _1\right\rangle $, $\left| \Psi
_2\right\rangle $, $\cdots $, and $\left| \Psi _n\right\rangle $ are $n$
linearly-independent states of a system A, there exist a unitary operator $U$
and a measurement $M$, which together yield the following evolution 
\begin{equation}
\label{1}\left| \Psi _i\right\rangle \left| \Sigma \right\rangle \stackrel{%
U+M}{\longrightarrow }\left| \Psi _i\right\rangle \left| \Psi
_i\right\rangle ,\text{ }\left( i=1,2,\cdots ,n\right) ,
\end{equation}
where $\left| \Sigma \right\rangle $ is the input state of a system B.
System A and B each has an $N$-dimensional Hilbert space with $N\geq n$.

To prove the above statement, we introduce a probe P with an $n_p-$%
dimensional Hilbert space, where $n_p\geq n+1$. Suppose $\left|
P_0\right\rangle $, $\left| P_1\right\rangle $, $\cdots $, and $\left|
P_n\right\rangle $ are $n+1$ orthogonal states of the probe P. If there
exists a unitary operator $U$ to make 
\begin{equation}
\label{2}U\left( \left| \Psi _i\right\rangle \left| \Sigma \right\rangle
\left| P_0\right\rangle \right) =\sqrt{\eta }\left| \Psi _i\right\rangle
\left| \Psi _i\right\rangle \left| P_0\right\rangle +\stackrel{n}{%
\stackunder{j=1}{\sum }}c_{ij}\left| \Phi _{AB}^{\left( j\right)
}\right\rangle \left| P_j\right\rangle ,
\end{equation}
where $\left| \Phi _{AB}^{\left( 1\right) }\right\rangle $, $\left| \Phi
_{AB}^{\left( 2\right) }\right\rangle $, $\cdots $, and $\left| \Phi
_{AB}^{\left( n\right) }\right\rangle $ are $n$ normalized states of the
composite system AB (not generally orthogonal), in succession we measure the
probe P and the output state is preserved if the measurement result is $P_0$%
. With a probability $\eta $ of success, this measurement projects the
composite system AB into the replicated state $\left| \Psi _i\right\rangle
\left| \Psi _i\right\rangle $, where $i=0$, $1$, $\cdots ,$ or $n$.
Therefore, the evolution (1) exists if Eq. (2) holds. To prove existence of
the unitary operator $U$ described by Eq. (2), we first introduce two lemmas.

{\it Lemma 1.} If $\left| \phi _1\right\rangle $, $\left| \phi
_2\right\rangle $, $\cdots $, and $\left| \phi _n\right\rangle $ are $n$
orthonormal states, and $\left| \widetilde{\phi }_1\right\rangle $, $\left| 
\widetilde{\phi }_2\right\rangle $, $\cdots $, and $\left| \widetilde{\phi }%
_n\right\rangle $ are other $n$ orthonormal states, there exists a unitary
operator $U$ to make 
\begin{equation}
\label{3}U\left| \phi _i\right\rangle =\left| \widetilde{\phi }%
_i\right\rangle ,\text{ }\left( i=1,2,\cdots ,n\right) .
\end{equation}

{\it Proof.} Suppose the considered system has an $N$-dimensional Hilbert
space $H$ with $N\geq n$. There exist $N-n$ orthonormal states $\left| \phi
_{n+1}\right\rangle $, $\left| \phi _{n+2}\right\rangle $, $\cdots $, and $%
\left| \phi _N\right\rangle $, which together with $\left| \phi
_1\right\rangle $, $\left| \phi _2\right\rangle $, $\cdots $, and $\left|
\phi _n\right\rangle $ make an orthonormal basis for the space $H$.
Similarly, the states $\left| \widetilde{\phi }_1\right\rangle $, $\left| 
\widetilde{\phi }_2\right\rangle $, $\cdots $, $\left| \widetilde{\phi }%
_n\right\rangle $, $\left| \widetilde{\phi }_{n+1}\right\rangle $, $\cdots $
and $\left| \widetilde{\phi }_N\right\rangle $ make another orthonormal
basis. The following operator
\begin{equation}
\label{4}U=\stackrel{n}{\stackunder{j=1}{\sum }}\left| \widetilde{\phi }%
_j\right\rangle \left\langle \phi _j\right| 
\end{equation}
is unitary, which can be easily checked by verifying the identity $%
U^{+}U=UU^{+}=I$. The operator (4) yields the evolution (3). Lemma 1 is thus
proved.

{\it Lemma 2.} If $2n$ states $\left| \phi _1\right\rangle $, $\left| \phi
_2\right\rangle $, $\cdots $, $\left| \phi _n\right\rangle ,$ $\left| 
\widetilde{\phi }_1\right\rangle $, $\left| \widetilde{\phi }_2\right\rangle 
$, $\cdots $, and $\left| \widetilde{\phi }_n\right\rangle $ satisfy 
\begin{equation}
\label{5}\left\langle \phi _i|\phi _j\right\rangle =\left\langle \widetilde{%
\phi }_i|\widetilde{\phi }_j\right\rangle ,\text{ }\left( i=1,2,\cdots ,n;%
\text{ }j=1,2,\cdots ,n\right) , 
\end{equation}
there exists a unitary operator $U$ to make $U\left| \phi _i\right\rangle
=\left| \widetilde{\phi }_i\right\rangle ,$ $\left( i=1,2,\cdots ,n\right) $.

{\it Proof.} Suppose $\gamma _1=\left\| \left| \phi _1\right\rangle \right\| 
$, where the norm $\left\| \left| \phi \right\rangle \right\| $ is defined
by $\left\| \left| \phi \right\rangle \right\| =\sqrt{\left\langle \phi
|\phi \right\rangle }$. Let $\left| \phi _1^{^{\prime }}\right\rangle =\frac
1{\gamma _1}\left| \phi _1\right\rangle $ and 
\begin{equation}
\label{6}\left| \phi _j^{^{\prime }}\right\rangle =\frac 1{\gamma _j}\left[
\left| \phi _j\right\rangle -\stackrel{j-1}{\stackunder{k=1}{\sum }}%
\left\langle \phi _k^{^{\prime }}|\phi _j\right\rangle \left| \phi
_k^{^{\prime }}\right\rangle \right] ,\text{ }\left( j=2,3,\cdots ,n\right) 
\end{equation}
where $\gamma _j=\left\| \left| \phi _j\right\rangle -\stackrel{j-1}{%
\stackunder{k=1}{\sum }}\left\langle \phi _k^{^{\prime }}|\phi
_j\right\rangle \left| \phi _k^{^{\prime }}\right\rangle \right\| $. The $n$
states $\left| \phi _1^{^{\prime }}\right\rangle $, $\left| \phi
_2^{^{\prime }}\right\rangle $, $\cdots $, $\left| \phi _n^{^{\prime
}}\right\rangle $ are obviously orthonormal. On the other hand, following
Eq. (5), the $n$ states $\left| \widetilde{\phi }_1^{^{\prime
}}\right\rangle =\frac 1{\gamma _1}\left| \widetilde{\phi }_1\right\rangle $
and 
\begin{equation}
\label{7}\left| \widetilde{\phi }_j^{^{\prime }}\right\rangle =\frac
1{\gamma _j}\left[ \left| \widetilde{\phi }_j\right\rangle -\stackrel{j-1}{%
\stackunder{k=1}{\sum }}\left\langle \phi _k^{^{\prime }}|\phi
_j\right\rangle \left| \widetilde{\phi }_k^{^{\prime }}\right\rangle \right]
,\text{ }\left( j=2,3,\cdots ,n\right) 
\end{equation}
are also orthonormal. Hence, from the lemma 1, there exists a unitary
operator $U$ to make 
\begin{equation}
\label{8}U\left| \phi _i^{^{\prime }}\right\rangle =\left| \widetilde{\phi }%
_i^{^{\prime }}\right\rangle ,\text{ }\left( i=1,2,\cdots ,n\right) .
\end{equation}
Eq. (8) is just another expression of the evolution $U\left| \phi
_i\right\rangle =\left| \widetilde{\phi }_i\right\rangle $. Lemma 2 is thus
proved.

Now we return to the proof of the main theorem. If Eq. (2) holds, its $%
n\times n$ inter-inner-products yield the following matrix equation 
\begin{equation}
\label{9}X^{\left( 1\right) }=\eta X^{\left( 2\right) }+CC^{+},
\end{equation}
where the $n\times n$ matrixes $C=\left[ c_{ij}\right] $, $X^{\left(
1\right) }=\left[ x_{ij}^{\left( 1\right) }\right] =\left[ \left\langle \Psi
_i|\Psi _j\right\rangle \right] $, and $X^{\left( 2\right) }=\left[
x_{ij}^{\left( 2\right) }\right] =\left[ \left\langle \Psi _i|\Psi
_j\right\rangle ^2\right] $. Lemma 2 suggests, the converse of the above
statement also holds, i.e., if Eq. (9) is satisfied with a positive cloning
efficiency $\eta $, the unitary evolution (2) exists and hence the $n$
states $\left| \Psi _1\right\rangle $, $\left| \Psi _2\right\rangle $, $%
\cdots ,$ and $\left| \Psi _n\right\rangle $ can be cloned by the same
cloning machine.

To prove there is a positive $\eta $ to satisfy Eq. (9), first we show that
the matrix $X^{\left( 1\right) }$ is positive-definite. This is the
following lemma.

{\it Lemma 3.} If $n$ states $\left| \Psi _1\right\rangle $, $\left| \Psi
_2\right\rangle $, $\cdots ,$ and $\left| \Psi _n\right\rangle $ are
linearly-independent, the matrix $X^{\left( 1\right) }=\left[ \left\langle
\Psi _i|\Psi _j\right\rangle \right] $ is positive-definite.

{\it Proof.} For an arbitrary $n$-vector $B=\left( b_1,b_2,\cdots
,b_n\right) ^T$, the quadratic form $B^{+}X^{\left( 1\right) }B$ can be
expressed as 
\begin{equation}
\label{10}B^{+}X^{\left( 1\right) }B=\left\langle \Psi _T|\Psi
_T\right\rangle =\left\| \left| \Psi _T\right\rangle \right\| ^2,
\end{equation}
where 
\begin{equation}
\label{11}\left| \Psi _T\right\rangle =b_1\left| \Psi _1\right\rangle
+b_2\left| \Psi _2\right\rangle +\cdots +b_n\left| \Psi _n\right\rangle .
\end{equation}
If the states $\left| \Psi _1\right\rangle $, $\left| \Psi _2\right\rangle $%
, $\cdots ,$ and $\left| \Psi _n\right\rangle $ are linearly-independent,
the summation state $\left| \Psi _T\right\rangle $ is not zero for any $n$%
-vector $B$ and its norm is thus positive. Following Eq. (10), the matrix $%
X^{\left( 1\right) }$ is positive-definite. This proves the lemma 3.

Since $X^{\left( 1\right) }$ is positive-definite, from continuity, for a
small enough but positive $\eta $, the matrix $X^{\left( 1\right) }-\eta
X^{\left( 2\right) }$ is also positive-definite. So $X^{\left( 1\right)
}-\eta X^{\left( 2\right) }$ can be diagonalized by a unitary matrix $U_1$
as follows 
\begin{equation}
\label{12}U_1^{+}\left( X^{\left( 1\right) }-\eta X^{\left( 2\right)
}\right) U_1=diag\left( m_1,m_2,\cdots ,m_n\right) ,
\end{equation}
where all the eigenvalues $m_1,$ $m_2,$ $\cdots ,$ and $m_n$ are positive
real numbers. In Eq. (9), the superposition constants matrix $C$ can be
chosen as 
\begin{equation}
\label{13}C=U_1diag\left( m_1,m_2,\cdots ,m_n\right) U_1^{+}.
\end{equation}
Eq. (9) is hence satisfied with a positive cloning efficiency $\eta $. This
completes the proof of the main theorem.

In the above proof, the condition of linearly-independence of the $n$ states 
$\left| \Psi _1\right\rangle $, $\left| \Psi _2\right\rangle $, $\cdots ,$
and $\left| \Psi _n\right\rangle $ plays an essential role. If $\left| \Psi
_1\right\rangle $, $\left| \Psi _2\right\rangle $, $\cdots ,$ and $\left|
\Psi _n\right\rangle $ are linearly-dependent, there is an $n$-vector $B$ to
make $B^{+}X^{\left( 1\right) }B=0$ and the matrix $X^{\left( 1\right) }$ is
therefore only semi-positive-definite. With a positive cloning efficiency $%
\eta $, in general, $X^{\left( 1\right) }-\eta X^{\left( 2\right) }$ is no
longer a semi-positive-definite matrix. But the matrix $CC^{+}$ is
semi-positive--definite. So Eq. (9) can not be satisfied. This suggests in
an alternative way, $n$ linearly-dependent states $\left| \Psi
_1\right\rangle $, $\left| \Psi _2\right\rangle $, $\cdots ,$ and $\left|
\Psi _n\right\rangle $ can not be cloned by the same cloning  machine.

Theorem 1 can be extended. In fact, if $n$ states $\left| \Psi
_1\right\rangle $, $\left| \Psi _2\right\rangle $, $\cdots ,$ and $\left|
\Psi _n\right\rangle $ can be cloned, they can be multi-cloned. In the proof
of the theorem 1, the crucial point is, with a small enough cloning
efficiency $\eta $, the matrix $X^{\left( 1\right) }-\eta X^{\left( 2\right)
}$ is positive-definite for a positive-definite $X^{\left( 1\right) }$. This
results from the continuity. For a cloning machine yielding $n$-copies of
the input states, the matrix $X^{\left( 2\right) }$ correspondingly becomes $%
X^{\left( n\right) }==\left[ x_{ij}^{\left( n\right) }\right] =\left[
\left\langle \Psi _i|\Psi _j\right\rangle ^n\right] $. Similarly, $X^{\left(
1\right) }-\eta X^{\left( n\right) }$ is also a positive-definite matrix
with a small enough $\eta $. So $n$ linearly-independent quantum states can
be multi-cloned. This extends the theorem 1. Of course, to generate
multi-copies of the input states, the maximum cloning efficiency alters.

The proof of the theorem 1 is constructive, i.e., it gives a systematic
method for constructing the desired unitary evolution $U$ and the
measurement $M$. But in general, the Hamiltonian yielding the cloning is
very complicated. It remains an open question how to construct the simplest
cloning machine, which at the same time maximizes the cloning efficiency.

Quantum cloning has important applications in quantum cryptography. That
linearly-independent states can be cloned seemingly threatens the security
of the quantum cryptography schemes based on two non-orthogonal states
[3,4]. But this is not the case. The key reason is that, the cloning
efficiency $\eta $ can not attain $100\%$ for non-orthogonal input states.
If $\eta =1$, Eq. (9) reduces to 
\begin{equation}
\label{14}X^{\left( 1\right) }=X^{\left( 2\right) }.
\end{equation}
This is possible if and only if the $n$ input states $\left| \Psi
_1\right\rangle $, $\left| \Psi _2\right\rangle $, $\cdots ,$ and $\left|
\Psi _n\right\rangle $ are orthonormal or identical. So non-orthogonal
states can not be cloned with a $100\%$ efficiency. Sometimes, the
measurement of the probe does not yield the desired result $P_0$ and the
cloning fails. Through these failures, in quantum cryptography, the sender
and the receiver can find the intervention of the eavesdropper. Therefore,
in the ideal case ( in the noiseless channel ), quantum cryptography can not
be eavesdropped by cloning the transmitted states. However, in noisy
channels, quantum cloning may be adopted as an important eavesdropping mean.

Quantum cloning also has potential applications in quantum computation. In a
recent paper [14], Nielsen and Chuang described programmable quantum gate
arrays. It has been shown there a universal quantum gate array---a gate
array which can be programmed to perform any unitary operation on the
data---exists only if one allows the gate array to operate in a
probabilistic fashion. After a deterministic unitary evolution, joint
measurements should be made on the data and the program qubits to project
the state of the gate array onto some subspace with a certain probability of
success. The crucial question in quantum programming is to raise this
probability. Similarly, in the cloning here, non-orthogonal states are
replicated also in a probabilistic way and the main concern is to increase
the cloning efficiency, which corresponds to the probability of success in
quantum programming. More connections may be found in these two subjects.

At last, we remark that quantum cloning is different from the inaccurate
quantum copying. Inaccurate quantum copying recently arouses great interests
[15-18]. In the inaccurate copying, the evolution is limited to the unitary
process and the input-output state fidelity can not attain 1 for
non-orthogonal states. The input states, whether linearly-independent or
linearly-dependent, all can be inaccurately copied. However, in the cloning
here, through some unitary-reduction process, the input states are
accurately replicated, though the cloning efficiency can not attain 100\%
for non-orthogonal states. Only linearly-independent quantum states can be
cloned.\\

{\bf Acknowledgment}

This project was supported by the National Natural Science Foundation of
China.

\newpage\

\end{document}